# Novel Muon Tomography Detector for the Pyramids

**Richard T. Kouzes[1], Alain Bonneville[1,2], Azaree Lintereur[3], Isar Mostafanezhad[4], Ryan Pang[4], Ben Rotter[4], Farjana Snigdha[4], Michael Tytgat[5], Shereen Aly[6], Basma ElMahdy[7], Yasser Assran[8], and Ayman Mahrous[9]**

[1] Pacific Northwest National Laboratory, Richland, Washington 98119, USA
[2] Oregon State University, Corvallis, Oregon 97331, USA
[3] The Pennsylvania State University, State College, Pennsylvania 16801, USA
[4] Nalu Scientific, Honolulu, Hawaii 96822, USA
[5] Ghent University, Ghent, Belgium
[6] Helwan University and Canadian International College, Cairo, Egypt
[7] Helwan University and The British University in Egypt, Cairo, Egypt
[8] Suez University and The British University in Egypt, Suez, Egypt
[9] Egypt-Japan University of Science and Technology, Alexandria, Egypt

## Abstract

Cosmic-ray muons which impinge upon the Earth's surface can be used to image the density of geological and man-made materials located above a muon detector. The detectors used for these measurements must be capable of determining the muon rate as a function of the angle of incidence. Applications of this capability include geological carbon storage, natural gas storage, enhanced oil recovery, compressed air storage, oil and gas production, tunnel detection, and detection of hidden rooms in man-made structures such as the pyramids. For these applications the detector must be small, rugged, and have operational characteristics which enable its use in remote locations, such as low power requirements. A new muon detector design is now being constructed to make measurements on the Khafre pyramid, in Egypt, to look for unknown voids that might exist in the structure. The new detector design uses monolithic plates of scintillator with wavelength shifting fiber optic readout to obtain location information. This design will meet the operational requirements, while also providing a geometry which can be modified for different measurement conditions.

## 1. Introduction

Muon tomography is a method of imaging that utilizes cosmic ray produced muons to reveal density changes and structures in containers, buildings, or environments. The term "tomography" can imply a three-dimensional imaging method but is used here in an overarching manner to refer to muon imaging which may be either a two-dimensional projection or a three-dimensional reconstruction depending on the method utilized. Several fields require imaging measurements of such density changes in a subsurface or ground-level environment, including but not limited to, archaeology [1, 2], tunneling activities [3, 4], carbon sequestration [5, 6], non-proliferation and treaty verification [7, 8], nuclear smuggling [9, 10], oil and gas exploration and storage [3], mineral exploration [11], and volcano studies [12].

*Source of Muons*

The muon, discovered in 1936, is a lepton with a half-life of 1.56  s and a mass of 105.7 MeV/c$^2$. Muons with a broad range of energies, from MeV to TeV, are generated in the Earth's atmosphere at approximately 15 km above the surface by energetic cosmic rays (mostly protons) that produce pions which subsequently decay to a muon plus a neutrino. There are about 160 muons s$^{-1}$·m$^{-2}$ impinging on the Earth at sea level with an angular distribution that follows closely a cos$^2$ law with respect to the vertical [13]. At these energies, muons are minimally ionizing particles, meaning that they deposit approximately the same amount of energy in a given material per unit path length independent of the muon's energy; these muons lose about 2 MeV gm$^{-1}$·cm$^{-2}$ of energy in materials. This simplifies the design requirements of a muon detector. Since these muons are highly penetrating, with a roughly exponential fall off in the flux with depth in the Earth, they can be detected at reasonable rates to several hundred meters below ground.

Muons undergo small angle scattering and attenuation proportional to the atomic number and density of the material through which they move. This creates the opportunity to determine the density of the material that a muon traverses before it reaches a detector.

*Muon Detection*

Muon detectors are typically constructed from gas proportional tubes, plastic scintillator, resistive plates, or multiwire chambers. If a muon detector can determine the x-y coordinate of a muon at two (vertically) separated positions, the direction of the muon can be determined. Muon tomography thus uses two or more muon detector planes to provide the trajectory of individual muons. By making measurements of many muons, deviations from the cos$^2$ distribution can be observed, potentially indicating the presence of voids or dense objects as a function of direction above the detector planes. The time required to perform statistically significant measurement depends on the depth of the detector (overburden), the detector size and efficiency, and the size of the void or dense inclusion to be detected and is typically hours to days. Fortunately, muon detectors normally have an intrinsic efficiency that is close to 100%, and background is strongly rejected through coincidence.

For muon tomography, the detector consists of two or more planes of detector material, which determine the x-y coordinate of a muon's passage. This is shown conceptually in Figure 1, where the red line is the transiting muon, the green lines represent readout positions in the x direction, while the y readout positions are on the underside of the planes.

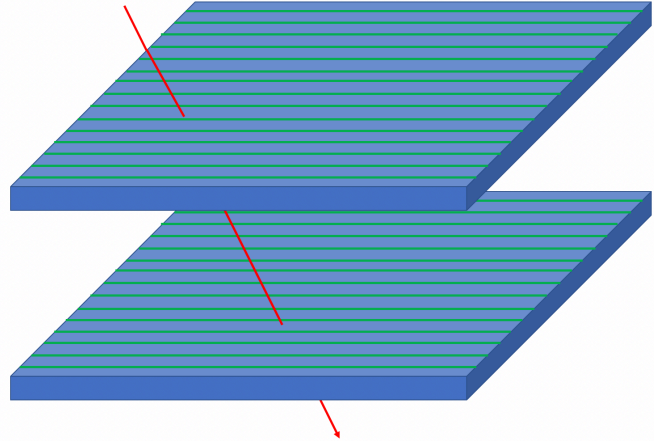

**FIGURE 1.** Detector concept example with two x-y layers.

The angular resolution of the detector is determined by the precision of the x and y position measurements in each plane and the spacing distance between the planes. There are many examples in the literature describing such muon tomography detector systems of varying size and composition [1, 2, 3, 4, 5, 6, 7, 8, 9, 10, 11].

Figure 2 shows a concept for a muon tomography detector small enough (15 cm x 60 cm) to fit into a 6 to 8-inch borehole consisting of a pair of x-y detection planes made from polystyrene scintillator rods [14, 15]. Figure 3**Error! Reference source not found.** shows the prototype of this muon detector built at Pacific Northwest National Laboratory to be used for borehole measurements, where the white rods are the two pairs of layers of polystyrene scintillator separated by a gap. Each rod of scintillator generates light which is transported by a (green) wavelength-shifting fiber optic to the end of the rod where the light was then observed by silicon photomultipliers (SiPMs). Such a detector design has the advantage that it can be built in any size by expanding the size of the planes (number of rods), does not need any gas flow, is mechanically robust, and only uses low voltages for the SiPM detectors. This detector was designed for application to a carbon sequestration project to look for density changes above the detector due to the insertion of carbon dioxide into a storage reservoir.

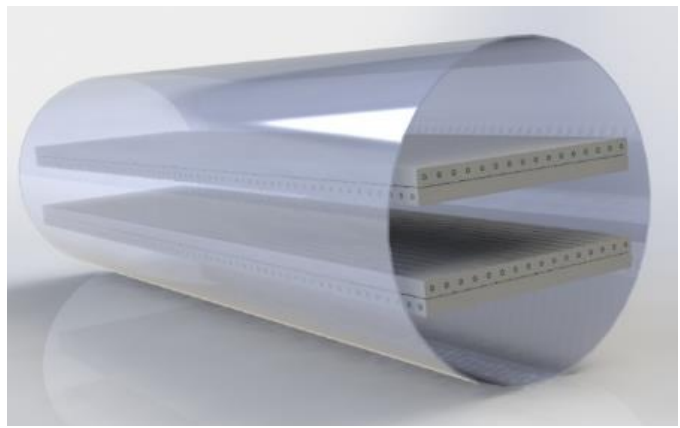

**FIGURE 2.** Concept for a muon tomography detector designed to fit into a borehole**.**

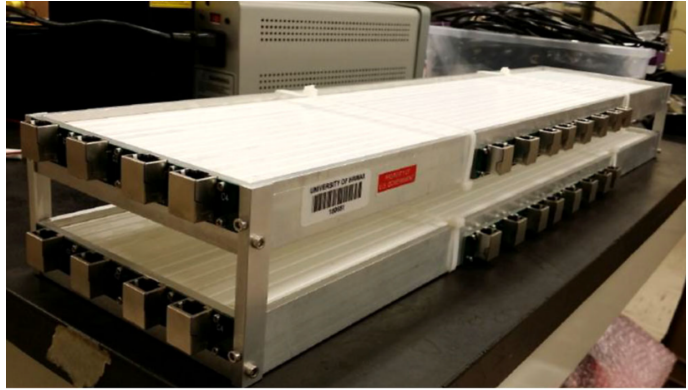

**FIGURE 3.** Small muon detector showing the two detection planes [5].

## 2. Application Examples

One possible application for a muon detector is to search for hidden rooms within the great pyramids of Giza. If located in the bottom of a pyramid, a muon detector can be used to search for cavities above the detector. Searching for tunnels would be similar, where the muon detector is below the void being investigated. These applications compare simulation results to measurements to look for deviations in the attenuation of the muon flux due to unknown voids above the detector. For pyramid applications, the ability to position the detector in a small underground space with the ability to adjust the angular resolution is critical to identifying small voids above. Since this is a closed space, a scintillator detector has a significant safety advantage over gas detectors which use asphyxiating gases.

A second application example is for monitoring spent nuclear fuel in a storage cask where the high-density uranium in the fuel causes muon scattering, allowing the imaging of the fuel position and the ability to detect any changes in the fuel load. The ability to position the detector around the cask is essential to the measurement. The muon detector can be left in place for extended periods (years) to monitor for any changes.

A third example is an application to homeland security. Decision Sciences (San Diego, CA) designed a large muon tomography system to screen trucks at border crossings for contraband material. This system takes a few minutes to acquire enough muon scattering data to detect the presence of high-density objects like special nuclear material.

A fourth example is the carbon sequestration application of the detector mentioned above, where a series of the borehole muon detectors would be placed under a sequestration site where they would monitor the change in density in the storage reservoir due to carbon dioxide displacing water in the rock. Monitoring the signal over long time periods would indicate changes in the amount of stored $CO_2$.

## 3. Novel Detector Concept

Following the development of the carbon sequestration borehole detector shown in Figure 3, a concept for a new detector design for measurements in the pyramids was developed based on a solid block of scintillator rather than individual rods. The solid plate of polyvinyl toluene (PVT) plastic scintillator has grooves milled into both surfaces in orthogonal directions on the two sides, as shown in Figure 4. As seen in Figure 5**Error! Reference source not found.**, when a muon (red line) traverses the PVT (blue), it produces scintillation light, some of which travels to fiber optics (green) in multiple grooves in the PVT plate. The fiber optics transport the light to SiPMs at the ends of each fiber. By collecting the light from multiple fibers in each plane, the position of the muon transit can be interpolated to more precision than the fiber spacing of 1 cm.

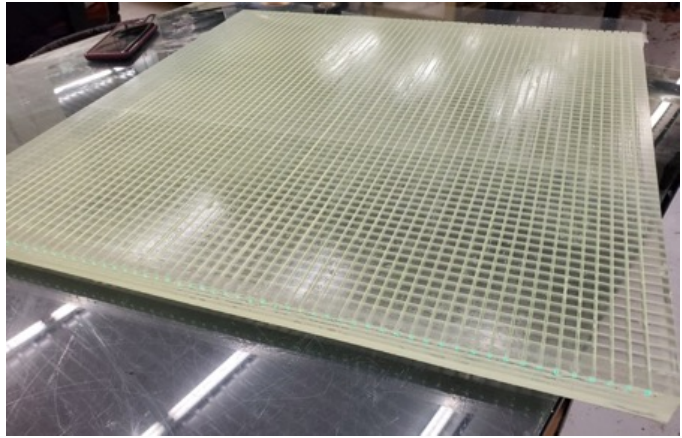

**FIGURE 4.** Crossed fiber readout of PVT scintillator plate.

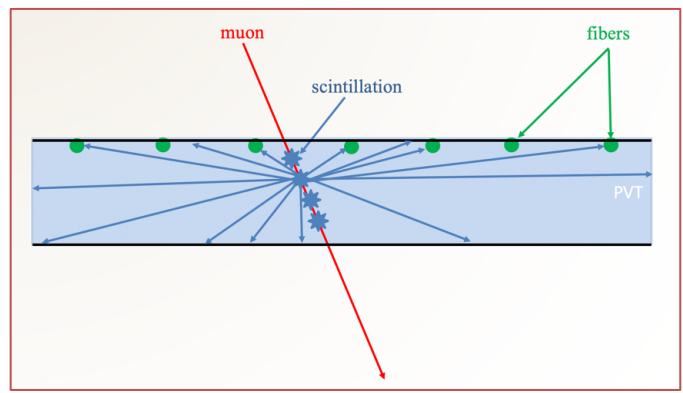

**FIGURE 5.** A muon deposits energy in the PVT, producing light that is picked up by fibers.

This detector is reconfigurable, consisting of two such x-y detection planes housed in separate boxes that can be spaced at a variable distance in order to obtain adjustable angular resolution. The two detector planes determine the muon trajectory. Thus, they can be used to look for scattering above the pair of panels, such as for the pyramid application, or placed on two sides an object of interest to look for scattering within the object, such as a fuel storage cask. The size of the detector plates, spacing and configuration of the detector is determined by the application.

A big advantage of this new design is it eliminates the complexity of previous designs that use multiple detection components (rods of scintillator or gas tubes) which require a mechanical structure to maintain them in alignment, and may have limitations on configurability (e.g., maximum spacing of the planes). In this new design, the two plates of plastic scintillator, each of dimension 61 cm square, are positioned with an adjustable separation (nominally 20 cm to 200 cm). The plates are of appropriate thickness (nominally 2 cm) to provide the required position resolution (better than 5 mm). This design is more robust and simpler to construct than previous designs since it consists of only two sections. This design allows angular resolution that is adjustable by changing the separation between the plates, with each of the PVT plates housed in individual light-tight boxes.

The readout of the scintillator consists of longitudinal embedded wavelength shifting fibers in the surface of both sides of each PVT scintillator plate, as shown in Figure 4. The fibers on each side are in orthogonal directions to allow determination of the x-y position of a muons passage. These fibers allow four-fold coincidence for background rejection, and reconstruction of the incidence muon angle. Each fiber is read out by a SiPM on the end of the fiber. A muon passing through the detector plates loses energy and thus generates scintillation light in the material. This light is collected by multiple fibers in proportion to their distance from the track location, allowing better localization through extrapolation based on the signal size. The SiPMs are mounted on printed circuit boards mounted on the scintillator plate edge and the signals are sent by cable to digitizer electronics (CAEN model DT5550W plus A55PET4 PETIROC4 boards, shown in Figure 6).

The data is read out from the CAEN instrument using a laptop computer for analysis. Coincidence is required between multiple fiber signals between planes during data acquisition and during analysis in order to reject background events. Integrated electronics with measurement of temperature provides consistent data over time. Analysis of the signal from the multiple fibers allows determination of the location of the muon passage through the detector, resulting in computation of the directional vector components for each muon.

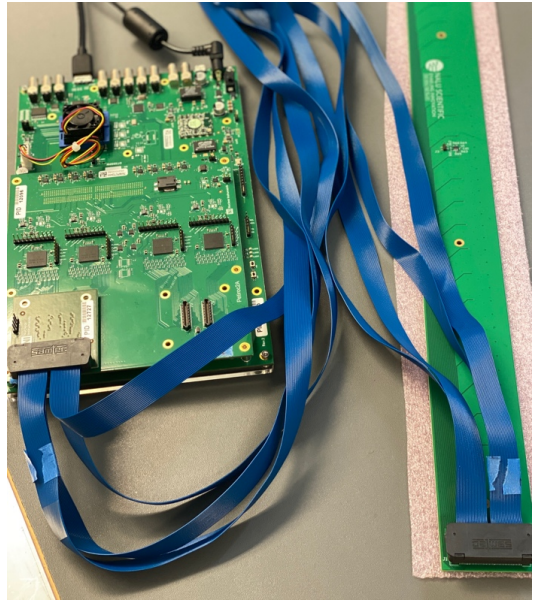

**FIGURE 6.** CAEN DT5550 readout connected to the muon board with SiPMs.

## 4. Conclusions

The borehole muon detector demonstrated that the approach of using scintillating rods with fiber optic wavelength shifting fibers and SiPM readout could be successfully implemented [16]. The new detector design based on PVT plates with wavelength shifting fibers and SiPM readout for application to searching for voids in the pyramids in 2022 is expected to be equally successful in providing the needed angular resolution to discover or rule out voids in these ancient structures.

## Acknowledgements

This detector is being developed through fund provided by the Egyptian Academy of Scientific Research and Technology (ASRT), Project ID: 6379. PNNL is operated for the US Department of Energy by Battelle under contract DE-AC05-76RLO 1830. PNNL-SA-167973. No conflict of interest exists.

## References

[1] K. Morishima et al., Nature 552, 386 (2017).

[2] H. Gómez. Nuclear Inst. and Methods in Physics Research, A 936, 14 (2019).

[3] R. Kaiser. Phil. Trans. R. Soc. A 377, 20180049 (2018).

[4] E. Guardincerri, C. Rowe, E. Schultz-Fellenz, M. Roy, N. George, C. Morris, J. Bacon, M. Durham, D. Morley, K. Plaud-Ramos, D. Poulson, D. Baker, A. Bonneville, R. Kouzes. Pure and Applied Geophysics 174, 2133 (2017).

[5] A. Bonneville, R. Kouzes, J. Yamaoka, A. Lintereur, J. Flygare, G. Varner, I. Mostafanezhad, R. Mellors, E. Guardincerri, C. Rowe. Phil. Trans. R. Soc. A 377, 20180060 (2018).

[6] V.A. Kudryavtsev, N.J.C. Spooner, J. Gluyas, C. Fung, M. Coleman, International Journal of Greenhouse Gas Control 11, 21 (2012).

[7] H.M. O'D. Parker, M.J. Joyce. Progress in Nuclear Energy 85, 297e318 (2015).

[8] S. Chatzidakis, R. Howard, H. Gadey, A. Farsoni. ORNL/SPR-2020/1728, September 30, 2020.

[9] G. Bonomi, P. Checchia, M. D'Errico, D. Pagano, G. Saracino. Progress in Particle and Nuclear Physics 112, 103768 (2020).

[10] P. Checchia. JINST 11 C12072 (2016).

[11] D. Varga, G. Nyitrai, G. Hamar, G. Galgóczi, L. Oláh, H.K.M. Tanaka, T. Ohminato. Nuclear Inst. and Methods in Physics Research, A 958, 162236 (2020).

[12] H. K.M. Tanaka, et al., Earth and Planetary Science Letters, *263*, 104 (2007).

[13] C. Hagmann, D. Lange, D. Wright. IEEE Nuclear Science Symposium Conference record (2007).

[14] A. Bonneville, J. Flygare, R. Kouzes, A. Lintereur, G. Varner, J. Yamaoka. IEEE Transactions on Nuclear Science 65, 2724 (2015).

[15] J. Flygare, A. Bonneville, R. Kouzes, J. Yamaoka, A. Lintereur. IEEE Transactions on Nuclear Science 65, 2724 (2018).

[16] A. Bonneville, R. Kouzes, J. Yamaoka, C. Rowe, E. Guardincerri, J.M. Durham, C.L. Morris, D.C. Poulson, K. Plaud-Ramos, D.J. Morley, J.D. Bacon, J. Bynes, J. Cercillieux, C. Ketter, K. Le, I. Mostafanezhad, G. Varner, J. Flygare, A.T. Lintereur. Nuclear Inst. and Methods in Physics Research A 851, 108 (2017).